\documentclass{mem}
\usepackage{natbib}\usepackage{txfonts}\usepackage{balance}
\usepackage{graphicx}
\usepackage[a4paper,breaklinks,dvipdfm]{hyperref}
\idline{75}{282}
\begin{document}
\def\teff{$T\rm_{eff }$}
\def\kms{$\mathrm {km s}^{-1}$}

\title{
AGN astrophysics from comparing radio and Gaia optical astrometry
}

\subtitle{Relativistic jets and gravitational wave rockets}

\author{Ian \,Browne}

  \offprints{Ian  Browne}

\institute{Jodrell Bank Centre for Astrophysics,\\ 
University of Manchester, \\
Oxford Road, \\
Manchester M13 9PL, U.K.}

\authorrunning{Browne }

\titlerunning{Displaced AGN}

\abstract{Gaia will open up a huge volume of new parameter space in
which to explore the physics of AGN and black hole evolution. We
address the question as to how far along the relativistic jets
blazar radio, optical and gamma ray emission originated. In some
models the optical centroid wander should be detectable as the relative
contributions of thermal and non-thermal optical emission
change. Black holes powering AGN do not necessarily reside at the
centres of their host galaxies; they can be one member of a binary
pair or they could have received a kick after binary
coalescence. In radio-loud AGN comparison of astrometric radio and
and optical positions can reveal such displacements. It is suggested
that it would be feasible to do this using Gaia and e-MERLIN for a 
sample of thousands of elliptical radio galaxies. 

}
\maketitle{}

\section{Introduction}
There are numerous ways in which the combination of Gaia optical
astrometry and accurate radio astrometry can be used to investigate
important astrophysical questions. Several of these are covered in
detail in other contributions to the meeting. I will focus primarily
on two questions: the physics of radio jets, in particular where does
the radio emission arise with respect to the central black hole, and
searching for AGN displaced from the centres of their host galaxies,
either because there is a binary black hole or the black hole has
received a large velocity kick as a result of the coalescence of two
black holes. Though there may be recognizable electromagnetic
signatures for kicked black holes (see \cite{kom12} for a recent review),
astrometry offers the only possibility for unambiguous detections.



\section{The astrophysics of radio jets}

We start by summarizing what we think we know about radio jets and then
discuss open questions which might be addressed astrometrically.

\begin{itemize}

\item Jets are produced by accreting black hole systems

\item They are emitted in opposite directions with relativistic speeds
having bulk Lorentz factors ranging up to 10 or 20 and perhaps more. For
this reason it is nearly always only the approaching jet that is visible

\item At all wavelengths jets flare on a range of timescales of
weeks to many years

\item Using VLBI, superluminal motion of blobs in the jets is detected
and the emergence of new moving blobs appear to be related to flaring activity.

\item The radio, infrared, optical, and sometimes X-ray emission, is
produced by the synchrotron mechanism. The jets have a secondary
peak in their spectral energy distributions which is usually
attributed to inverse Compton emission. Surveys of the high latitude sky are
dominated by the same objects at gamma-ray frequencies as at GHz
frequencies.

\item The optical and gamma ray emission seems to come from the same
part of the jet. This conclusion us based on evidence that there is a
correlation with no appreciable time delay \citep{wag11} between
optical and gamma-ray flare activity. Similarly, comparison of
multi-epoch VLBI data and gamma-ray variability suggest that in many
cases the radio and gamma-ray emission regions are co-spatial.

\end{itemize}

We can be reasonable confident of most of the above. Astrometry is
unlikely to have direct impact on the really big issue of the
mechanism of jet production. There are, however, important secondary
questions concerning where things happen that are certainly in the
realm of astrometric investigation. The answers to these should give indirect
pointers to jet production physics. We would like to know from where the
non-thermal jet emission at all wavelengths originates relative to the
thermal AGN emission the latter being the best marker we have for the position
of the black hole.

There is an active debate concerning the location all the non-thermal
emission. On the one hand \cite{mar11}, \cite{agu11} and
\cite{jor11} present convincing observational evidence for this
emission occurring $\sim$10~pc or more downstream from the jet
origin. On the other hand \cite{ster11} present strong theoretical
arguments for the gamma-ray emission originating within the BLR; i.e
$\sim$0.1 ~pc from the jet base. Which of these scenarios is correct,
or if both are correct in the sense the position of the emitting
regions is different in different types of object, has knock on
consequences. For example, the density of AGN thermal photons in the
non-thermal emitting region will depend on how far downstream the
latter is. This is important because there is a debate concerning the
relative importance of internally generated photons and external AGN
photons in the production of the gamma-ray emission \citep{gup11}.

How can astrometry help? At z = 1, a typical quasar redshift, 1~mas is
about 8~pc and hence well within the realm to be probed by Gaia
astrometry. Imagine a flat spectrum radio quasar (FSRQ) in which the
contributions to the total emission of the blazar and the thermal disk
are roughly equal. If the blazar emission originates $\sim$10~pc away
from the core and the blazar component varies, then the centroid of
the combined emission will shift up and down the jet. Thus finding
optical centroid shifts, especially of the position shifts correlated
with the brightness of an object would be strong support for the
scenario in which the non-thermal emission regions were well outside
the BLR (see contribution be Sonia Anton). Further support would be
forthcoming if the direction of shift was found to be along the jet
axis as determined from VLBI. More direct comparison of VLBI radio
positions and Gaia optical ones are covered in other
contributions. This kind of comparison can tell us directly if the jet
non-thermal optical and radio emission originate from the same place
but not where both occur relative to thermal emission arising from
accretion on to the black hole.

\section{Searching for displaced black holes}

\noindent The idea is to pick a large sample of passive elliptical
galaxies which have detectable compact radio emission and look for
displacements between the centroid of the stellar light as measured
with Gaia and the centroid of the radio emission. The basic premise is
that the stellar light will define the centre of mass and the radio
emission the position of the black hole. The radio astrometry could be
done with $\sim$1~mas precision using the e-MERLIN array in the
UK. Significant displacements could arise due to either a binary black
hole system with only one black hole powering a radio source or a ``kicked''
black hole after the coalescence of a binary black hole system 
before the merged black hole settles back to the centre of mass \citep{fat04}.

There is an extensive literature on the life cycle of black holes
during the hierarchical merging process that has built the galaxies we
see today (e.g. \cite{vol10}; \cite{bak08}). In a seminal paper by
\cite{beg80} they describe how after galaxies merge, their associated
nuclei which consist of a central stellar cluster and black hole,
rapidly merge and a binary black hole system forms in the new enlarged
stellar cluster. Dynamical friction shrinks the orbit on a timescale
of $\sim10^{8}$~yr at which point most of the stars that contribute
the dynamical friction are depleted and the orbit can ``stall'' at a
radius $\leq$1~pc. How long this stalled phase lasts is very uncertain
but is very relevant to the redshift when most black hole coalescences
occur. If it is the order of a Hubble time many coalescences should be
happening at the present epoch but if much shorter most will have
happened at a much higher redshift near the epoch at which the galaxy
merger rate peaks.

Though the \cite{beg80} analysis gives good description of the
period up to black hole coalascence it is only in the last few years
that numerical general relativity has been able to treat the actual
coalescence phase. The prediction of the emission of a burst of
gravitational radiation is expected and well known. What is new and
remarkable is that under the right initial conditions the
single black hole produced in the process receives a large kick in
velocity, occasionally greater than the escape velocity of the
galaxy, but often large enough to be detected (\citep{fat04}.

Both binaries and kicked black holes could give rise to detectable
displacements of an active nucleus from the centre of mass of the host
galaxy. There are many imponderables influencing how likely one is to
detect measurable displacements. Various characteristic timescales are
important:

\begin{enumerate}

\item The length of time that it takes for a pair of black holes to
reach stalling orbit after their parent galaxies merge. It is only during this
pre-stalled phase that the separation will be large enough to be detected.

\item The length of time the orbit remains stalled. If this is short
most coalescences will happen at high redshift when the merger rate
is greatest.

\item If a kick is given when the black holes coalesce, how long is it
before the black hole settles down to the centre of mass.

\end{enumerate}

The primary motivation for a search for binary and kicked black holes
is to try and pin down these times and timescales. The time after host
galaxy merger and the associated black holes reaching stalling radius
is expected to be $\sim 10^{8}$~yr, thus perhaps a few percent of high
redshift galaxies may host binary black holes with orbital radii
$\geq$1~pc. In how many there might be detectable radio emission from
at least one of the black holes is very uncertain though the existence
of two detectable radio nuclei in 0402+379 with a separation of 7.3~pc
\citep{rod06} is encouraging. On the other hand in a search of 3114
sources with VLBI maps was made by \cite{bur11}. She was looking for
multiple nuclei with separations $\leq$100~mas and 0402+379 was the only
binary black hole candidate detected. Thus the occurrence of two orbiting
black holes both of which are radio loud is rare.

No double radio nuclei with wider separations were found in the
$\sim$ 16000 200~mas-resolution radio maps made during the CLASS
gravitational lens search \citep{brow03}. On the other hand we are
talking about the probability of both black holes being active radio
sources. If we, for example, take it that the statistics indicate that
perhaps one in 10,000 radio sources have two radio nuclei, then, a
perhaps reasonable guess would be that 1 in 100 might have one of the
pair radio loud. The fundamental problem in the case of a single radio
loud nucleus is finding a position reference with
respect to which the black hole can be measured. The position
reference problem can be solved in principal for passive elliptical
galaxies because the starlight can be used as a proxy to estimate the
position of the centre of mass (see below).

Estimating the likelihood of being able to detect kicked black holes is
probably even more uncertain than that for binary black holes. The
stalling time before coalescence is not well constrained. However, it
is believed that after coalescence the timescale for the kick hole to
settle back to the centre of the merged galaxy is also $\sim
10^{8}$~yr. But we do not know if the black hole will be capable of
powering a radio source during this time. Simulations have been
performed by \cite{sij11} focused on the amount of optical AGN
emission expected but not of the expected radio emission. Furthermore,
depending upon how long the orbits remain stalled before final
coalescence will determine whether there are any black holes remaining
``kicked'' in the relatively low redshift universe.

Passive elliptical radio galaxies (i.e. having no optical evidence for
an AGN) form the best targets. The elliptical galaxy light should
define the galaxy position and the compact radio nucleus defines the
position of the black hole. There are still some significant
uncertainties. We do not know if kicked black holes will be (radio)
active. There are, though, encouraging hints; there is a claim for a
displacement detected in M87 \citep{bat10}.

Another practical uncertainty is how well measuring the centroid of
the visible light defines the centre of mass. Tidal tails, twisting
isophotes and faint companion galaxies could all cause problems. A
pilot study should be done. Because real black hole displacements will
be the exception rather than the rule a sample of a few tens of radio
ellipticals with the best available ground-based astrometry should be
picked and the positions of their radio cores measured. One would hope
that the spread in radio-optical position differences would be
consistent with the known astrometric errors. If the results were
encouraging this could lead on to a major programme to measure several
thousand radio positions with e-MERLIN for comparison with Gaia
optical positions. In a similar programme to the one outlined above
\cite{con11} have been using a combination of VLBA observations for
the radio astrometry and 2MASS for the optical astrometry and find
that the displacements are consistent with the 2MASS astrometric
errors. Clearly Gaia will be have much better astrometric accuracy
than 2MASS.

\section{Conclusions}

\noindent Two areas where Gaia can have a major impact in questions of
major importance in extragalactic astrophysics are in the physics of
radio jets in blazars and on elucidating the life histories of black
holes during and after galaxy mergers. In the case of blazars one
would look for time-variable displacements of the radio and optical
centroids and see if these displacements were correlated with
variations in optical total intensity. If detected this would be good
evidence for the jet emission arising $\geq$ a few parsec from the
galactic nucleus.

Predictions of the hierarchical merging picture of galaxy formation are
that black hole binaries should form, after a time they should
coalesce and occasionally the black hole produced by the coalescence
should receive a velocity kick large enough to produce a displacement
of $\sim$kpc from the centre of mass of the host galaxy. The timescales
for many of these stages in the life histories of supermassive black
holes are uncertain and the only way they can be constrained is by
quantifying the number of binary black holes and trying to detect
kicked black holes. Comparing the radio positions of compact sources
with the Gaia-measured optical centroids of elliptical galaxies offers
an unique opportunity to get a handle on some of these illusive
numbers.

\begin{acknowledgements}
I am grateful to Sonia Anton and Simon Garrington for numerous useful
discussions
\end{acknowledgements}

\bibliographystyle{aa}

\end{document}